\documentclass[,final,numberedheadings]{aipproc}

\layoutstyle{8x11single}

\begin{document}

\title{HFB calculations with a microscopic pairing interaction}

\author{T. Duguet\footnote{Present address: National Superconducting Cyclotron Laboratory, 1 Cyclotron Laboratory, East-Lansing, MI 48823; E-mail~: duguet@nscl.msu.edu}}{
  address={ Argonne National Laboratory, Physics Division, 9700 South Cass Avenue, Argonne, IL 60439, USA}
}

\author{P. Bonche}{
  address={Service de Physique Th\'eorique, CEA Saclay, 91191 Gif sur Yvette Cedex, France}
}

\begin{abstract}
Hartree-Fock-Bogolyubov (HFB) calculations making use of a recently proposed microscopic effective pairing interaction are presented. The interaction was shown to reproduce the pairing properties provided by the realistic $AV18$ force very accurately in infinite matter. Although finite-ranged and non-local, it makes 3D HFB calculations in coordinate space tractable. As a first application, basic pairing properties of calcium isotopes in their ground-state are studied. By comparing the results with those obtained using a standard Density-Dependent Delta Interaction, the crucial isovector character of the microscopic interaction is highlighted.
\end{abstract}

\maketitle

\section{Introduction}
\label{intro}

The structure of the nucleus and the properties of extended nuclear systems strongly depend on their possible superfluid nature. In finite nuclei, pairing constitutes the main part of the residual interaction and has a strong influence on most of low-energy properties of the system~\cite{bender03b}. In extended systems such as neutron stars, pairing is a decisive ingredient of dynamical and thermal evolutions~\cite{sauls,heiselberg3}. Despite its major role, the present knowledge of the pairing force and the nature of pairing correlations in nuclei, that is, the way Cooper pairs are formed in the nuclear medium out of the strong nucleon-nucleon ($NN$) force, is quite poor. Properties such as the range of the effective pairing interaction, its link to the bare force, its possible surface character in finite nuclei and its density dependence (in particular isovector) still have to be clarified~\cite{doba3,duguet2,doba4,bulgac3}.

Regarding self-consistent mean-field calculations of finite nuclei, only phenomenological forces such as the Gogny force~\cite{decharge80a} or (Density-Dependent) Delta Interactions ((DD)DI)~\cite{duguet2,doba4,bertsch,rigol} have been used in the pairing channel so far. One exception exists however~\cite{barranco04a}, where the Bardeen-Cooper-Schrieffer (BCS) gap equation was solved in the Hartree-Fock (HF) basis of $^{120}Sn$ using realistic $AV18$ bare force~\cite{wiringa}. It required the treatment of single-particle states up to $800$ MeV! Although successful in describing low-energy nuclear structure over the {\it known} mass table~\cite{bender03b}, the Gogny force and DDDI lack a clear link to the bare nucleon-nucleon ($NN$) interaction\footnote{Even if some significant differences remains, one can argue that the Gogny force behaves almost like a bare force in the $^{1}S_0$ and $^{1}D_2$ channels, especially when the D1S~\cite{d1s} parameterization is used~\cite{kucharek2,duguet04a}.}. This feature strongly limits the reliability of their analytical structure such as their possible density dependence. Also, their direct fit to nuclear data through mean-field calculations makes probable the re-normalization of beyond-mean-field effects. This is a significant limitation if one wants to go explicitly beyond that level of approximation. Finally, their fits performed onto very limited sets of nuclei around stability make their extrapolated use toward the drip-lines unsafe. For instance, while such phenomenological forces all provide similar and reasonable pairing properties around stability, the predicted location of the two-neutron drip-line can differ by up to ten/twenty mass units depending on the force used~\cite{bennaceur03a}.

To improve on that situation, a microscopic effective interaction explicitly linked to the bare $NN$ force, and equivalent to it at the mean-field level, was proposed recently to treat pairing correlations in the $^{1}S_0$ channel~\cite{duguet04a}. BCS pairing properties provided in infinite matter by $AV18$ were reproduced very accurately. These properties dealt not only with the gap at the Fermi surface as a function of density, but also with the momentum dependence of the gap at fixed density~\cite{duguet04a}.

In the present paper, we discuss the first results of 3D HFB calculations using that force and we focus on the isovector and low-density properties of the interaction. In section~\ref{formalism}, we briefly outline the formalism and the characteristics of the new pairing force. Results obtained along the calcium isotopic chain are discussed in section~\ref{results}. They are compared to those obtained using a standard surface peaked DDDI and a zero-range approximation to the new microscopic force. Conclusions are given in section~\ref{conclusions}.

\section{Formalism}
\label{formalism}

To treat pairing, one needs to specify the many-body technique used and the appropriate interaction to insert into the calculation at the chosen level of approximation. The latter depends both on the situation and on the system. In the present case, we concentrate on a (self-consistent) mean-field description of finite nuclei using the HFB method. Eventually, calculations beyond the mean-field are to be performed. Typically, correlations associated with symmetry restorations (Projected Mean Field Method) and large amplitude motion (Generator Coordinate Method) are considered~\cite{ring80a,bender03c,duguet03c}. Thus, one has to identify the appropriate vertices to be used coherently at each level of approximation.

While variational calculations are of no direct help in that respect, perturbative methods using Green's function or Goldstone formalisms provide guides to do so. In particular, such many-body theories show unambiguously that the interaction to be used in the particle-particle (p-p) channel at lowest order in irreducible vertices is the bare $NN$ force\footnote{This constitutes our motivation for the mean-field level. With the $G$-matrix~\cite{brueckner54a} in the p-h channel, such a mean-field theory precisely aims at treating the nucleus as a system of independent {\it pairs}, including the correlations associated with the existence of Cooper bound-states in the medium.}. At the next order, the irreducible pairing vertex involves the so-called polarization diagrams. This situation is in contrast to the particle-hole (p-h) channel where one may need to regularize the repulsive core of the bare interaction from the outset through the definition of an in-medium two-body vertex like the $G$-matrix~\cite{brueckner54a}. This stresses the fact that the effective forces may differ in the two channels at a given level of approximation.

The approximate ground-state energy $E^{HFB}$ of a nucleus is a functional of the one-body density matrix $\rho^{q}_{ji} = \langle \, \Phi \, | \, c_{i}^{\dagger} \, c_{j} \, | \, \Phi \rangle$ and pairing tensor $\kappa^{q}_{jl} = \langle \, \Phi \, | \, c_{l} \, c_{j} \, | \, \Phi \rangle$, where $| \, \Phi \rangle \,$ is the HFB state\footnote{The quantities are defined in an arbitrary basis and the isospin quantum number $q$ ($1/2$ for neutrons and $-1/2$ for protons) is specified to make clear that isospin mixing is considered neither in the p-h channel nor in the p-p channel.}. For a general presentation of the HFB formalism, we refer to Ref.~\cite{ring80a}. Also, the two-basis method used to solve the HFB problem iteratively is discussed in Ref.~\cite{gall}. Finally, a detailed presentation of the method applied to the new interaction will be soon available~\cite{duguet05a}. 

The pairing field $\Delta^{q}_{ij}$ reads as:

\begin{equation}
\Delta^{q}_{ij} \, = \, \frac{\partial \, E^{HFB}}{\partial \, \kappa^{q \, \ast}_{ij}} \, = \, \frac{1}{2} \sum_{kl} \, \overline{\left(V^{\, ^{1}S_0} \right)}_{ijkl} \, \, \kappa^{q}_{kl} \, \, \, \, ,  \label{E31}
\end{equation}
where, as already explained, $V^{\, ^{1}S_0}$ is the bare $NN$ force. In Ref.~\cite{duguet04a}, the {\it effective} vertex was then introduced by recasting the previous gap equation written in the canonical basis\footnote{The canonical basis corresponds to the single-particle basis diagonalizing the one-body density matrix $\rho^{q}_{ji} = \rho^{q}_{i} \, \delta_{ji} = \rho^{q}_{\bar{i}}$ and putting the pairing tensor in its canonical form $\kappa^{q}_{kl} = \kappa^{q}_{k\bar{k}} \, \delta_{l\bar{k}}$. $\rho^{q}_{m}$ plays in the canonical basis the role of the usual BCS occupation number. Which states $(i,\bar{i})$ are paired is a by-product of the Bogolyubov transformation solution of the problem. States are not paired a priori as in the BCS approximation.} into a {\it fully equivalent} expression~\cite{duguet04a}:

\begin{equation}
\Delta^{q}_{i} \, \equiv \, \Delta^{q}_{\bar{i}i} \, = \, - \, \frac{1}{2} \, \sum_{m} \, \overline{\left(V^{\, ^{1}S_0} \right)}_{i\bar{i}m\bar{m}} \, \, \frac{\Delta^{q}_{m}}{2 \, E^{q}_{m}} \, = \, - \, \frac{1}{2} \, \sum_{m} \, 2 \, \rho^{q}_{m} \, \overline{\left({\cal D}^{\, ^{1}S_0^{2q}} (0)\right)}_{i\bar{i}m\bar{m}} \, \, \frac{\Delta^{q}_{m}}{2 \, E^{q}_{m}}   \, \, \, \, ,
\label{gapequationcanon}
\end{equation}
where ${\cal D}^{\, ^{1}S_0^{2q}} (0)$ is an off-shell in-medium two-body matrix summing p-p and h-h ladders in the superfluid phase. In Eq.~\ref{gapequationcanon}, the expression of the pairing tensor in its canonical form $\kappa^{q}_{m\bar{m}} = \Delta^{q}_{m} / 2 E^{q}_{m}$ was used, where $E^{q}_{m} = E^{q}_{\bar{m}} = \sqrt{(h^{q}_{m}-\mu^{q})^{2}+\Delta^{q \, 2}_{m}}$, while $h^{q}_{m}$ is the diagonal matrix element of the HF field in the canonical basis and $\mu^{q}$ the chemical potential. Identifying the two sides of Eq.~\ref{gapequationcanon}, the effective pairing vertex is naturally defined through its antisymmetrized matrix elements in the canonical basis of the Bogolyubov transformation as:

\begin{equation}
\overline{\left(V^{\, ^{1}S_0^{2q}}_{eff}\right)}_{i\bar{i}m\bar{m}} \, = \,  2 \, \rho^{q}_{m}  \, \, \, \overline{\left({\cal D}^{\, ^{1}S_0^{2q}} (0)\right)}_{i\bar{i}m\bar{m}}   \, \, \, \, .
\label{effectiveforce1}
\end{equation}

In Ref.~\cite{duguet04a}, ${\cal D}^{\, ^{1}S_0^{2q}} (0)$ was calculated explicitly in infinite matter starting from a separable form of the bare $NN$ interaction in the $^{1}S_0$ channel. It was studied in detail and shown to take a closed form in coordinate space:

\begin{equation}
{\cal D}^{\, ^{1}S_0^{2q}} (0) \left(\vec{r}_{1}, \, \vec{r}_{2}, \,  \vec{r}_{3}, \, \vec{r}_{4} \,\right) \, = \, \lambda^{\, ^{1}S_0} \, \,  \frac{1 - P_{\sigma}}{2} \, \, \int d \vec{r} \, \, f \left[\rho_{q} (\vec{r} \,)\right] \, \, \frac{e^{-\sum_{i=1}^{4} \, |\vec{r}-\vec{r}_{i}|^{2} / 2 \alpha^{2}}}{(2\pi)^{6}\alpha^{12}} \, \, \, \, \, , 
\label{effectiveforce2}
\end{equation}
where $P_{\sigma}$ is the spin-exchange operator\footnote{The force acting only in the relative $S$-wave, the projection on the spin singlet corresponds to a simultaneous projection on the isospin triplet. The $T=1$ neutron-proton pairing ($T_{z} = 0$ component) is not considered here.}, while $\alpha = \sqrt{0.52} \,  fm$ and $\lambda^{\, ^{1}S_0} = -840 \, MeV.fm^{3}$ denote the range and the intensity of the force, respectively. No further adjustment is to be made in finite nuclei. The functional $f \left[\rho_{q} (\vec{r} \,)\right]$ incorporates the density dependence of the $T_{z} = 2q$ component of the effective interaction~\cite{duguet04a}. The density dependence stems from the re-summation of p-p and h-h ladders in the medium. The effective vertex is thus finite ranged, non local, total-momentum dependent and density dependent\footnote{Although the effective vertex breaks total-momentum and angular-momentum conservation of the interacting pair, as well as isospin symmetry, it does so in such a way that the energy functional itself remains invariant under rotation in isospin and real spaces, as well as under translation~\cite{duguet05a}.}. However, the computing cost of corresponding 3D HFB calculations is, through the two-basis method~\cite{gall,terasaki4}, of the same order as for a zero-range interaction.

By re-summing the effect of pairs scattered at high-energy into the effective vertex, the latter is soft even if the bare force has a hard core. Also, a smooth cut-off $2 \, \rho^{q}_{m}$ emerges naturally in the gap equation through its recast. This cut-off further limits the necessity to use large basis sets as in Ref.~\cite{barranco04a} and makes zero-range approximations of the effective vertex meaningful. The pairing problem is regularized in a similar way to what was proposed in Refs.~\cite{bulgac3,bulgac2,bulgac1} using re-normalization techniques. It is worth noting that the derived cut-off differs from all ad-hoc ones used in connection with usual DDDI. Thus, a zero-range (ZR) approximation providing identical gaps at the Fermi surface in infinite matter was defined in Ref.~\cite{duguet04a}. The coefficients entering the functional $f \left[\rho_{q} (\vec{r} \,)\right]$ differ from the ones used in the finite range case. Performing such an approximation, the roles of the range and of the density dependence of the interaction could be disentangled~\cite{duguet04a}. In particular, the surface-enhanced character of phenomenologically optimized DDDI~\cite{duguet2,doba4} was demonstrated and shown to be, to a large extent, a way of re-normalizing the range of the interaction. Also it was shown that usual DDDI miss the low-density behavior of the effective pairing force. 

Both the finite-range microscopic force and its zero-range approximation depend on the density $\rho_{q} (\vec{r} \,)$ of the interacting particles rather than on the total matter density. It is to be contrasted with usual DDDI which often take the form:

\begin{equation}
D^{^{1}S_0^{2q}} \left(\vec{r}_{1}, \, \vec{r}_{2}, \,  \vec{r}_{3}, \, \vec{r}_{4} \,\right) \, = \, \lambda^{\, ^{1}S_0} \, \,  \frac{1 - P_{\sigma}}{2} \, \, \left[1 - \frac{\rho_{0}(\vec{r}_{1} \,)}{\rho_{c}} \, \right] \, \, \delta (\vec{r}_{1}-\vec{r}_{3}) \, \, \delta (\vec{r}_{2}-\vec{r}_{4}) \, \, \delta (\vec{r}_{1}-\vec{r}_{2})  \, \, \, \, \, ,
\label{DDDI}
\end{equation}
where $\rho_{0}(\vec{r})$ denotes the total matter density (the local scalar-isoscalar part of the one-body density-matrix) while $\rho_{c}$ is equal to (one-half) the saturation density for the surface (half-surface) type pairing force.

The role of the finite range, the isovector density-dependence and the low density behavior of the pairing force, as well as the regularization scheme used together with contact approximations has to be addressed in detail. While we briefly focus on the isovector properties of the interaction in the present communication, we refer to Ref.~\cite{duguet05a} for an extensive study of all other issues.

\section{Results}
\label{results}

We performed 3D HFB calculations of Ca ground-states from proton to neutron drip-lines~\cite{gall}. Good particle-numbers were approximately restored before variation through the Lipkin-Nogami (LN) procedure~\cite{nogami64a}. Self-consistent blocking and time reversal symmetry breaking were included in the calculations of odd isotopes. The Sly4 Skyrme force~\cite{chabanat} was used in the p-h channel. Each calculation was repeated three times using the microscopic finite range force defined through Eq.~\ref{effectiveforce2}, its zero range approximation and a standard surface peaked DDDI as given by Eq.~\ref{DDDI}~\cite{rigol}. The latter was adjusted together with a phenomenological cut-off defining an active window ($\pm 5 MeV$) around the Fermi energy. 

By keeping the force in the p-h channel fixed, we probe the vertex used in the pairing channel, including the self-consistent coupling between the two channels. Of course, properties of the force in the p-h channel have an impact on the results. In that respect, it is worth noting that the considered DDDI was adjusted on properties of (non-exotic) nuclei using the SLy4 parameterization in the p-h channel~\cite{rigol}, and thus, is consistent with the isoscalar effective mass $m^{\ast}/m = 0.7 $ predicted by the latter. The microscopic pairing forces were adjusted once for all without any reference to finite nuclei with the property of not depending explicitly on the effective mass appearing in the p-h channel~\cite{duguet04a}. However, according to our definition of the mean-field, they should be used at the HFB level together with a p-h vertex providing an effective mass consistent with a $G$-matrix supplemented by a three-body force~\cite{zuo02b}.

\begin{figure}
  \includegraphics[angle=270,scale=0.28]{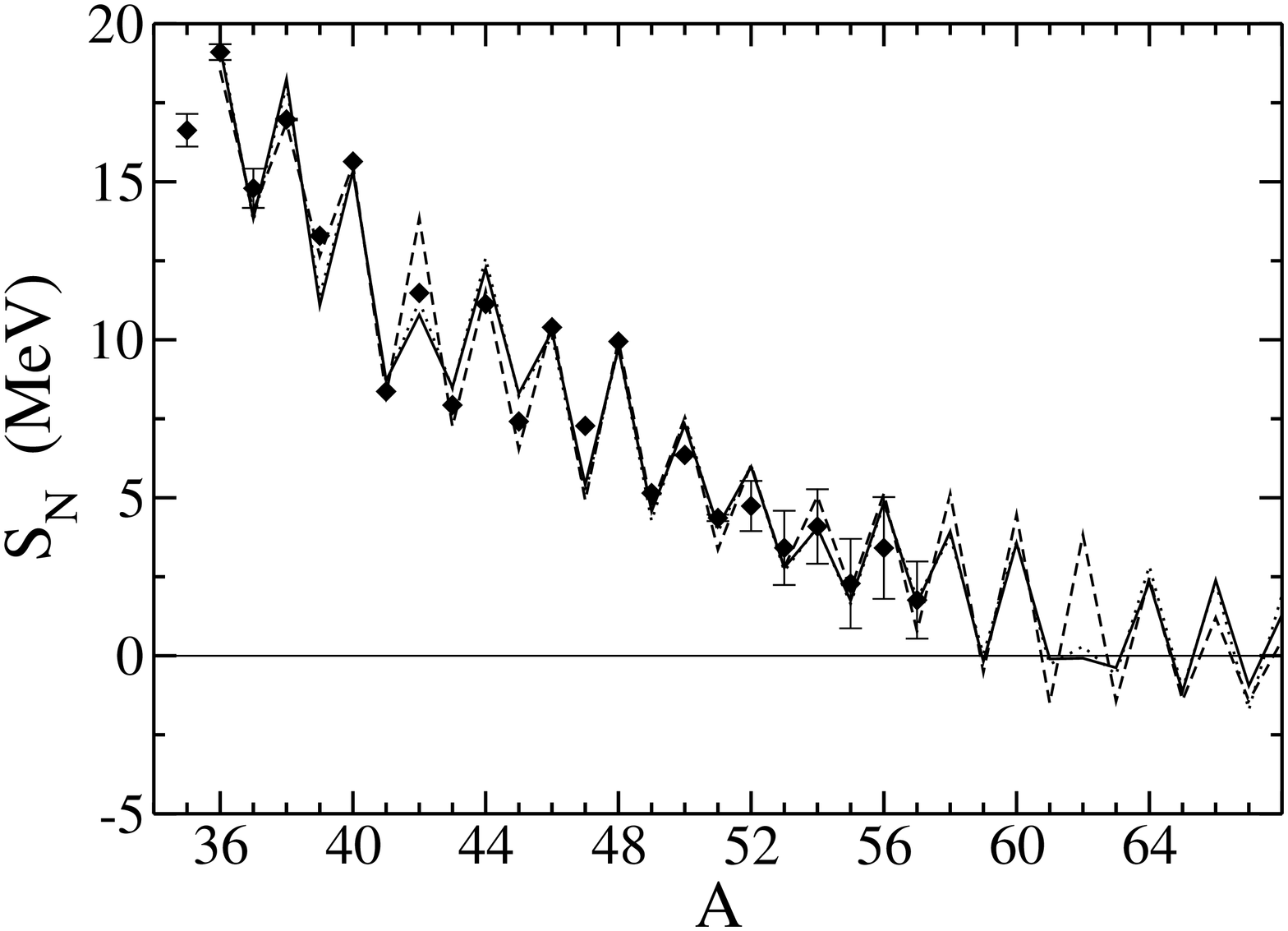} \hspace{-0.9cm} \includegraphics[angle=270,scale=0.28]{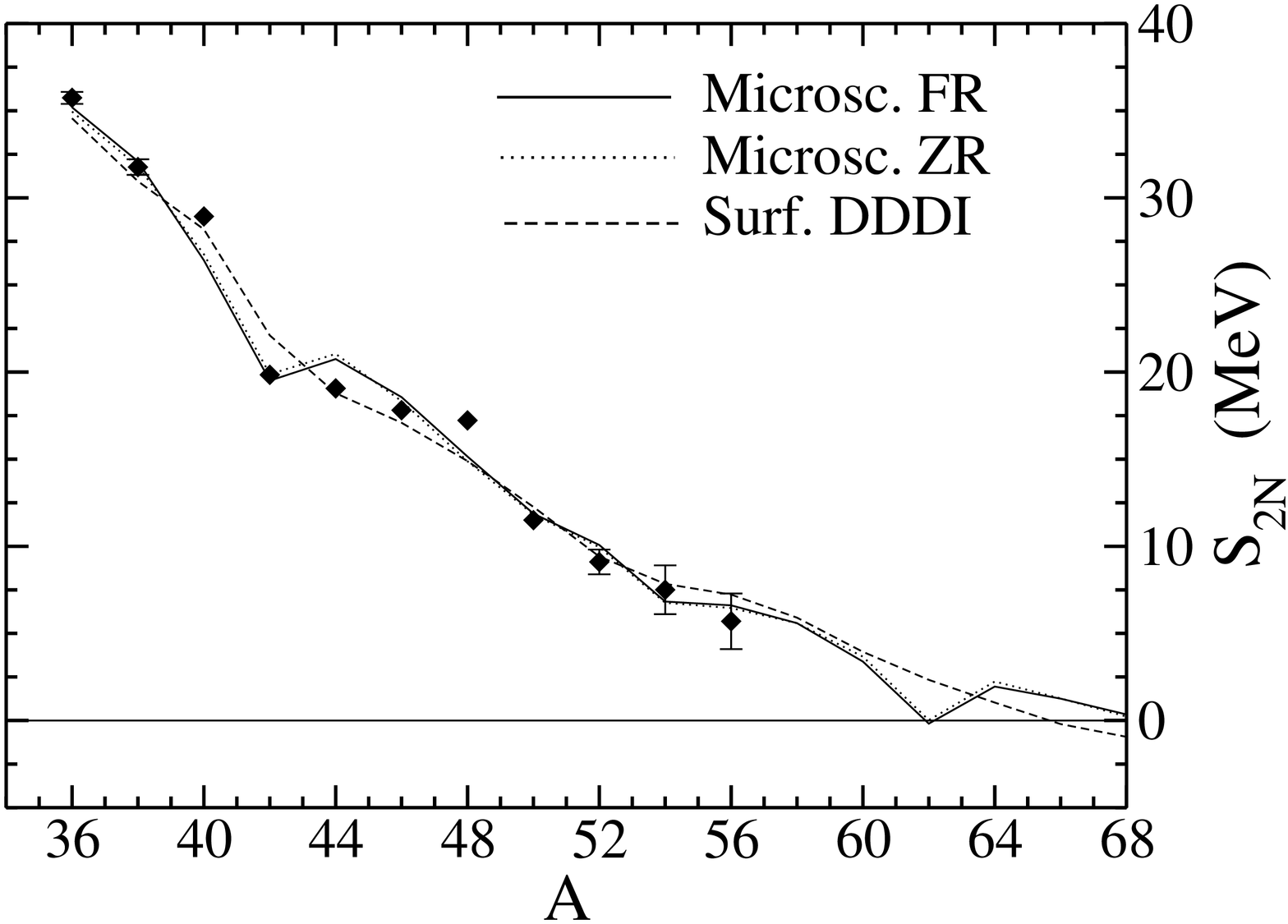}
  \caption{Left panel: one-neutron separation energy $S_{N} (A) = E (N-1,Z) - E (N,Z)$ in Ca isotopes. Right Panel: two-neutron separation energy $S_{2N} (A) = E (N-2,Z) - E (N,Z)$ for the same nuclei. Experimental~\cite{audi03a} (diamonds) and theoretical values for three different pairing forces are displayed.}
\label{sns2n}
\end{figure}

In Fig~\ref{sns2n}, calculated one and two-neutron separation energies are compared to experiment~\cite{audi03a}. Overall, the agreement is (at least) of the same quality as the one obtained from other calculations of the same type~\cite{bulgac3,bulgac4}. There are interesting differences which are beyond the scope of the present discussion. We rather concentrate on the effect of the pairing force used in the region of unknown exotic nuclei, because it is where its choice is of crucial importance. As an example, its influence on the position of the two-neutron drip-line is clearly seen on the right panel of Fig.~\ref{sns2n}. While the usual DDDI predicts it to be located at $N = 44$, the stability against two-neutron emission extends up to $N = 50$ when the microscopic forces are used. Note that a difference of two mass units in the predicted position of the drip-line for such light semi-magic nuclei can translate into a difference of ten mass units for lead isotopes~\cite{bennaceur03a}.

\begin{figure}
  \includegraphics[angle=270,scale=0.28]{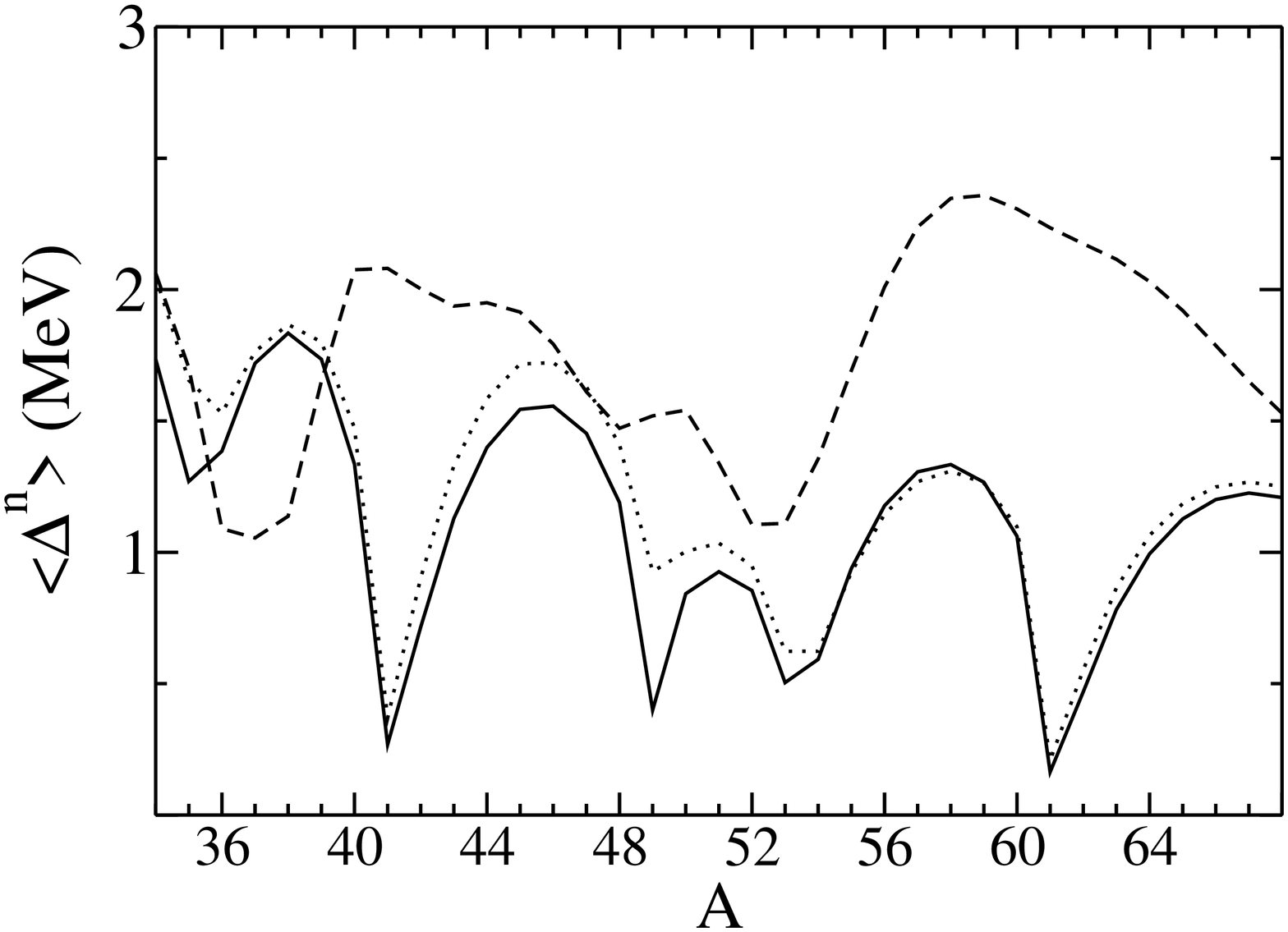} \hspace{-0.9cm} \includegraphics[angle=270,scale=0.28]{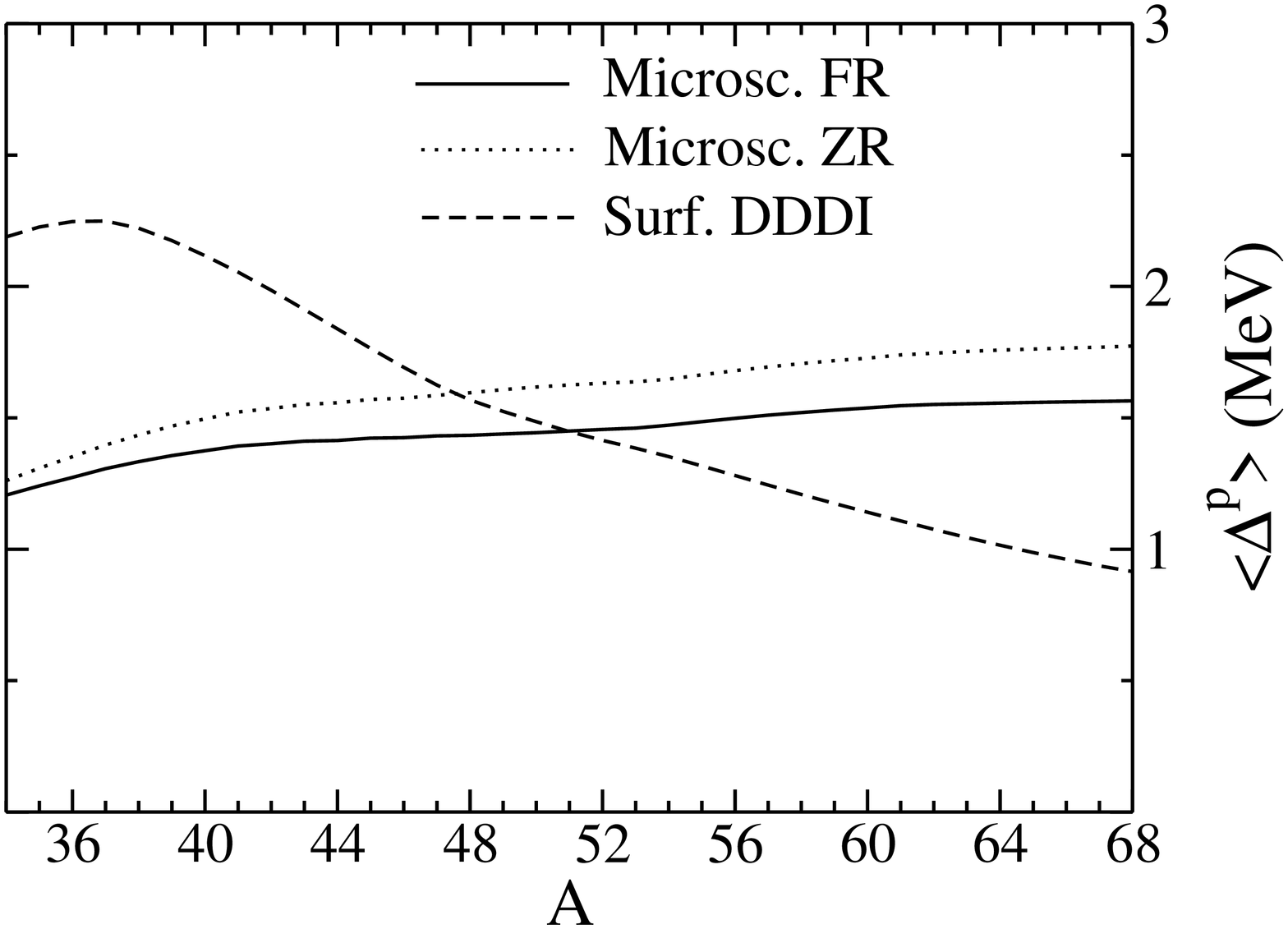}
  \caption{Neutron (left panel) and proton (right panel ) average pairing gaps $< \Delta^{q} (A) > = \sum_{n} \, \Delta^{q}_{n} \, \kappa^{q}_{n\bar{n}} \, /  \, \sum_{n} \kappa^{q}_{n\bar{n}}$ calculated along the Ca isotopic chain with three different pairing forces.}
\label{gapmoyens}
\end{figure}

Average neutron and proton pairing gaps calculated with the three pairing forces are plotted in Fig~\ref{gapmoyens} along the Ca isotopic chain\footnote{Values appear for odd and even particle numbers. For odd particle numbers, no blocking was considered here as such HFB states constitute the proper reference on top of which the blocking as to be eventually performed to describe the final odd state~\cite{duguet02a}. The average gaps are not compared with experiment because we do not want to discuss here how they are related to odd-even mass differences~\cite{duguet02b}.}. We see non-trivial differences between the predictions of the phenomenological DDDI and of the microscopic forces. We refer to Ref.~\cite{duguet05a} for a detailed discussion. Here we simply stress the different isovector trend of those predictions. While the magnitude of the neutron gaps are of the same order for nuclei around the stability line, the phenomenological DDDI provides much too strong pairing in neutron rich nuclei as anticipated in Ref.~\cite{duguet04a}. The overshoot of simple DI by usual DDDI near the drip-line was also identified~\cite{bennaceur03a} and often thought to be a result of the surface-peaked character of the latter\footnote{The way each interaction couples to the continuum also plays a role. Usually, pairing correlations are diminished by a strong coupling to the continuum. The way the continuum is treated numerically also influences the results near the drip-line~\cite{grasso01a}.}. In fact, in both cases, the primary cause of the overshoot is the improper isovector character of usual surface-enhanced DDDI associated with their dependence on the total (isoscalar) density\footnote{It is proved here not to be related to the range of the interaction. The gaps obtained with the finite range force do not differ significantly from those obtained with its zero-range approximation.}. Indeed, such a DDDI, when adjusted on nuclei having very similar neutron and proton densities, will provide stronger (weaker) pairing in a region of neutron rich matter than a force independent of the density or depending on the neutron (proton) density. This situation is also illustrated in the right panel of Fig.~\ref{gapmoyens}. While the LN prescription is responsible for the non-zero value of the proton gap at $Z=20$, the latter should not evolve much with neutron number. However, $<\Delta^{p} (A) >$ presents an artificial slope when using the DDDI whose origin goes precisely along the line of the previous argument. When repeating the calculation without LN, the proton gap even switches on artificially for $N \leq 24$ in the case of the DDDI.

\begin{figure}
  \includegraphics[angle=270,scale=0.28]{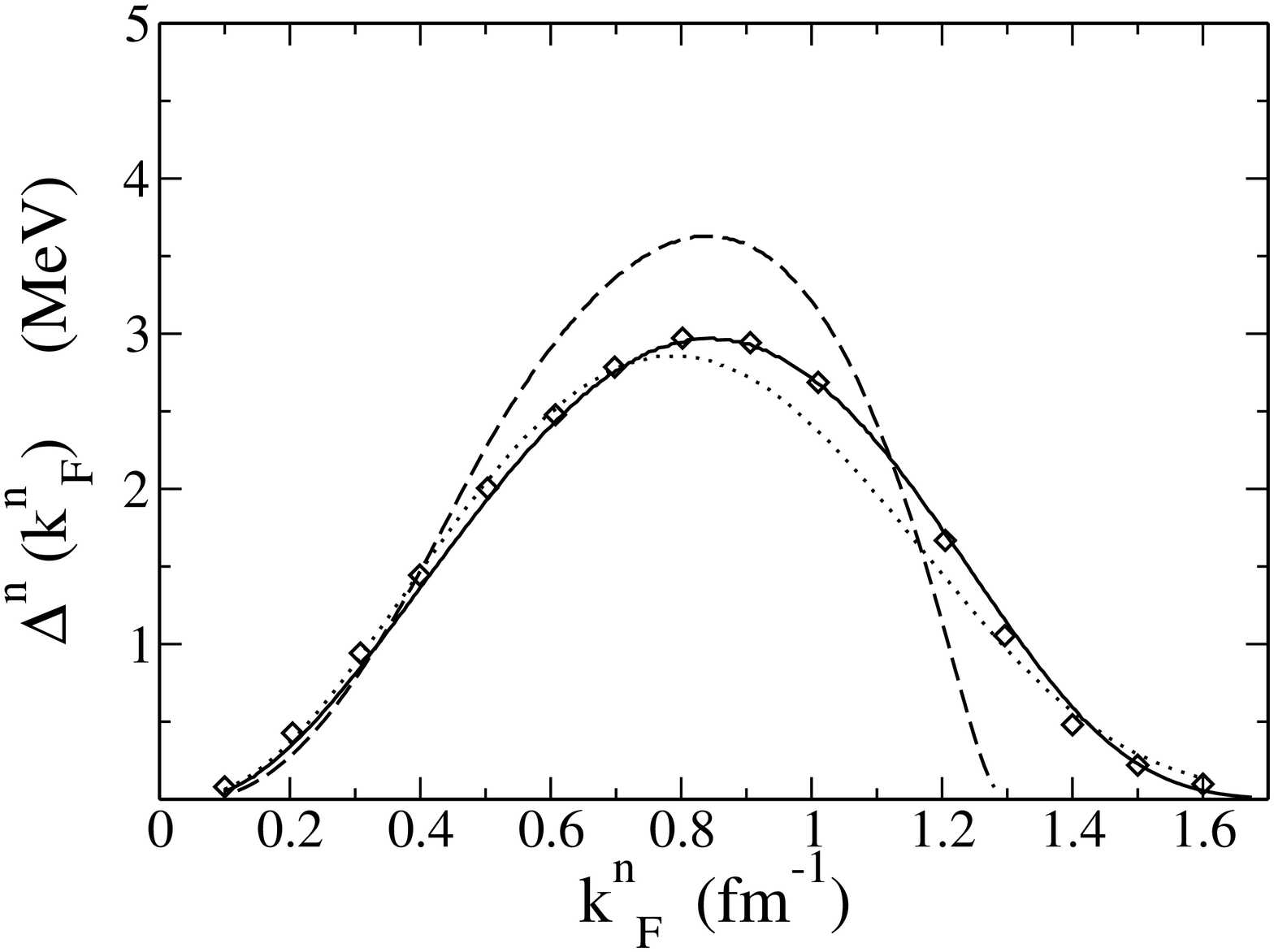} \hspace{-0.9cm} \includegraphics[angle=270,scale=0.28]{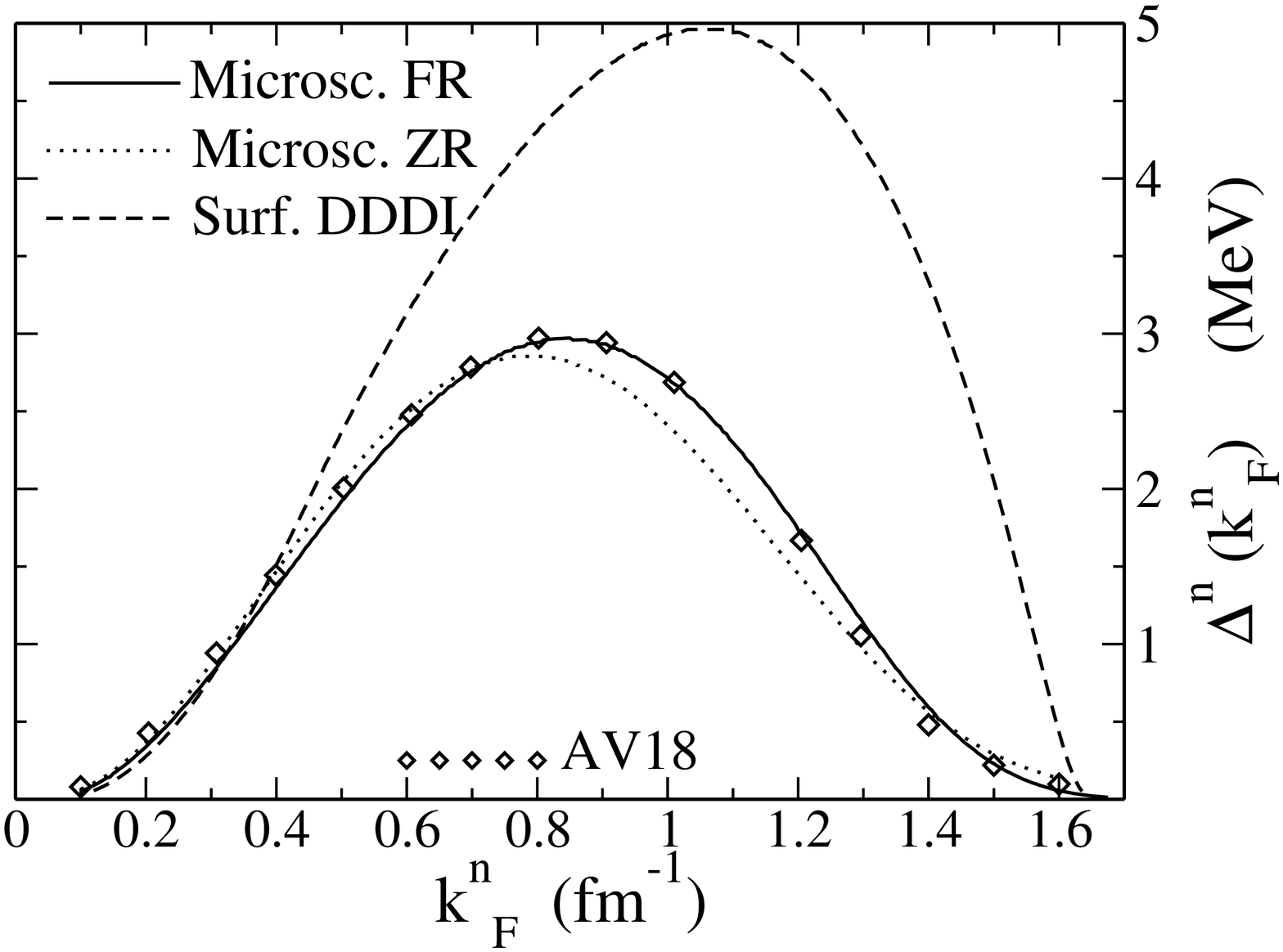}
  \caption{Neutron pairing gap at the (neutron) Fermi surface as a function of the neutron density. Left panel: symmetric nuclear matter. Right panel: pure neutron matter. The results obtained with the three pairing forces used in the present paper are compared to the results derived from the realistic AV18 $NN$ force~\cite{wiringa}. To make the theoretical comparison clear, all gaps are calculated with free kinetic energies as single-particle energies.}
  \label{figuregapmatinf}
\end{figure}

In order to confirm that interpretation, the BCS neutron gap at the Fermi surface in symmetric nuclear matter and in pure neutron matter is plotted in Fig.~\ref{figuregapmatinf} as a function of the neutron Fermi momentum $k^{n}_{F}$. Because we use free kinetic energies as single-particle energies, $\Delta^{n} (k^{n}_{F})$ obtained from AV18 or from the density-dependent microscopic forces are the same in symmetric matter and in pure neutron matter. On the other end, while the DDDI adjusted on stable nuclei reproduces rather well the gap in symmetric matter, it strongly overshoots it in neutron rich matter. This is due to its dependence on the total density and cannot be related here to any surface effect. Furthemore, beyond mean-field effects in the bulk one could eventually incorporate will always lead to a decrease of $\Delta^{n} (k^{n}_{F})$ in infinite neutron matter~\cite{shen03a}.

As a last point, we discuss the low-density behavior of the pairing force. It was shown in Ref.~\cite{duguet04a} that the intensity of the effective microscopic force strongly rises at low density due to the very large scattering length of the $NN$ interaction in the $^{1}S_{0}$ channel. The influence of this strong attraction can be seen in the right panel of Fig.~\ref{sns2n} where nuclei are stabilized beyond the sub-shell $N=40$ and the two-neutron drip-line pushed back by ten mass units. Indeed, the gaps obtained from the microscopic force resist their otherwise decreasing trend near the drip-line because of the increasing importance of low densities. Such an effect is not seen for the (still too strong) gaps calculated from the DDDI. Strong low-density dependence of DDDI were simulated phenomenologically in Ref.~\cite{doba4} and shown to bring about pathologies. This highlights the fact that such a density dependence should be used in connection with the corresponding microscopically derived cut-off.

\section{Conclusions}
\label{conclusions}

We presented results of the first (3D) HFB calculations performed in finite nuclei using a recently proposed microscopic effective pairing interaction. The isoscalar and isovector density-dependences derived ab-initio provide the pairing force with a strong predictive power when extrapolated toward the drip-lines. We concentrated here on that aspect by studying basic pairing properties of calcium isotopes in their ground-state. By comparing the results with those obtained from a standard Density-Dependent Delta Interaction, the crucial isovector character of the microscopic interaction was highlighted.

In the near future, more systematic HFB calculations will be presented to identify the role of the bare $NN$ force in building pairing in finite nuclei. Through comparisons with experiment, an indirect measure of missing correlations in the p-p channel will be realized. Local theories of pairing will be challenged by probing the importance of the finite range of the force, especially when describing low-energy excited states. Later, the use of the microscopic force will be extended to beyond mean-field calculations.

%%%%%%%%%%%%%%%%%%%%%%%%%%%%%%%%%%%%%%%%%%%%%%%%%%%%%%%%%%%%%%%%%%%%%%
%% You may have to change the BibTeX style below, depending on your %%
%% setup or preferences.					    %%
%%								    %%
%% If the bibliography is produced without BibTeX comment out the   %%
%% following lines and see the aipguide.pdf for further information %%
%%								    %%
%% For The AIP proceedings layouts use either			    %%
%%%%%%%%%%%%%%%%%%%%%%%%%%%%%%%%%%%%%%%%%%%%%%%%%%%%%%%%%%%%%%%%%%%%%%

\bibliographystyle{aipproc}   % if natbib is available

\bibliography{limits04}

\begin{thebibliography}{34}
\expandafter\ifx\csname natexlab\endcsname\relax\def\natexlab#1{#1}\fi
\providecommand{\enquote}[1]{``#1''}
\expandafter\ifx\csname url\endcsname\relax
  \def\url#1{\texttt{#1}}\fi
\expandafter\ifx\csname urlprefix\endcsname\relax\def\urlprefix{URL }\fi

\bibitem[Bender et~al.(2003)]{bender03b}
Bender, M., Heenen, P.-H., and Reinhard, P.-G., \emph{Rev. Mod. Phys.},
  \textbf{75}, 121 (2003).

\bibitem[Sauls(1989)]{sauls}
Sauls, J.~A., \emph{Timing Neutron Stars}, Dordrecht, Kluwer, 1989, p. 457.

\bibitem[Heiselberg and Hjorth-Jensen(2000)]{heiselberg3}
Heiselberg, H., and Hjorth-Jensen, M., \emph{Phys. Rep.}, \textbf{328}, 237
  (2000).

\bibitem[Dobaczewski and Nazarewicz(2003)]{doba3}
Dobaczewski, J., and Nazarewicz, W., \emph{Prog. Theor. Phys. Suppl.},
  \textbf{146}, 70 (2003).

\bibitem[Duguet et~al.(2001)]{duguet2}
Duguet, T., Bonche, P., and Heenen, P.-H., \emph{Nucl. Phys.}, \textbf{A679},
  427 (2001).

\bibitem[Dobaczewski et~al.(2001)]{doba4}
Dobaczewski, J., Nazarewicz, W., and Reinhard, P.~G., \emph{Nucl. Phys.},
  \textbf{A693}, 361 (2001).

\bibitem[Yu and Bulgac(2003{\natexlab{a}})]{bulgac3}
Yu, Y., and Bulgac, A., \emph{Phys. Rev. Lett.}, \textbf{90}, 222501
  (2003{\natexlab{a}}).

\bibitem[Decharg\'e and Gogny(1980)]{decharge80a}
Decharg\'e, J., and Gogny, D., \emph{Phys.\ Rev.}, \textbf{C21}, 1568 (1980).

\bibitem[Bertsch and Esbensen(1991)]{bertsch}
Bertsch, G.~F., and Esbensen, H., \emph{Ann. Phys. (NY)}, \textbf{209}, 327
  (1991).

\bibitem[Rigollet et~al.(1999)]{rigol}
Rigollet, C., Bonche, P., Flocard, H., and Heenen, P.-H., \emph{Phys. Rev.},
  \textbf{C59}, 3120 (1999).

\bibitem[Barranco et~al.(2004)]{barranco04a}
Barranco, F., Broglia, R., Colo', G., Vigezzi, E., and Bortignon, P.,
  \emph{Eur. Phys. J.}, \textbf{A21}, 57 (2004).

\bibitem[Wiringa et~al.(1995)]{wiringa}
Wiringa, R.~B., Stocks, V. G.~J., and Schiavilla, R., \emph{Phys. Rev.},
  \textbf{C51}, 38 (1995).

\bibitem[Berger et~al.(1991)]{d1s}
Berger, J.-F., Girod, M., and Gogny, D., \emph{Comp. Phys. Comm.}, \textbf{63},
  365 (1991).

\bibitem[Kucharek et~al.(1989)]{kucharek2}
Kucharek, H., Ring, P., and Schuck, P., \emph{Z. Phys.}, \textbf{A334}, 119
  (1989).

\bibitem[Duguet(2004)]{duguet04a}
Duguet, T., \emph{Phys.\ Rev.}, \textbf{C69}, 054317 (2004).

\bibitem[Bennaceur et~al.(2003)]{bennaceur03a}
Bennaceur, K., Bonche, P., and Meyer, J., \emph{C. R. Physique}, \textbf{4},
  555 (2003).

\bibitem[Ring and Schuck(1980)]{ring80a}
Ring, P., and Schuck, P., \emph{The Nuclear Many-Body Problem},
  Springer-Verlag, New-York, 1980.

\bibitem[Bender and Heenen(2003)]{bender03c}
Bender, M., and Heenen, P.-H., \emph{Nucl.\ Phys.}, \textbf{A713}, 390 (2003).

\bibitem[Duguet et~al.(2004{\natexlab{a}})]{duguet03c}
Duguet, T., Bender, M., Bonche, P., and Heenen, P.-H., \emph{Phys.\ Lett.},
  \textbf{B559}, 201 (2004{\natexlab{a}}).

\bibitem[Brueckner et~al.(1954)]{brueckner54a}
Brueckner, K.~A., Levinson, C.~A., and Mahmoud, H.~M., \emph{Phys. Rev.},
  \textbf{95}, 217 (1954).

\bibitem[Gall et~al.(1994)]{gall}
Gall, B., Bonche, P., Dobaczewski, J., Flocard, H., and Heenen, P.-H., \emph{Z.
  Phys.}, \textbf{A348}, 183 (1994).

\bibitem[Duguet et~al.(2004{\natexlab{b}})]{duguet05a}
Duguet, T., Bennaceur, K., and Bonche, P., unpublished (2004{\natexlab{b}}).

\bibitem[Terasaki et~al.(1995)]{terasaki4}
Terasaki, J., Heenen, P.-H., Bonche, P., Dobaczewski, J., and Flocard, H.,
  \emph{Nucl. Phys.}, \textbf{A593}, 1 (1995).

\bibitem[Yu and Bulgac(2002)]{bulgac2}
Yu, Y., and Bulgac, A., \emph{Phys. Rev. Lett.}, \textbf{88}, 042504 (2002).

\bibitem[Bulgac(2002)]{bulgac1}
Bulgac, A., \emph{Phys. Rev.}, \textbf{C65}, 051305 (2002).

\bibitem[Nogami(1964)]{nogami64a}
Nogami, Y., \emph{Phys. Rev.}, \textbf{134}, B313 (1964).

\bibitem[Chabanat et~al.(1997)]{chabanat}
Chabanat, E., Bonche, P., Haensel, P., Meyer, J., and Schaeffer, R.,
  \emph{Nucl. Phys.}, \textbf{A627}, 710 (1997).

\bibitem[Zuo et~al.(2002)]{zuo02b}
Zuo, W., Lejeune, A., Lombardo, U., and Mathiot, J., \emph{Eur. Phys. J.},
  \textbf{A14}, 469 (2002).

\bibitem[Audi et~al.(2003)]{audi03a}
Audi, G., H.Wapstra, A., and Thibault, C., \emph{Nucl. Phys.}, \textbf{A729},
  337 (2003).

\bibitem[Yu and Bulgac(2003{\natexlab{b}})]{bulgac4}
Yu, Y., and Bulgac, A., nucl-th/0302007 (2003{\natexlab{b}}).

\bibitem[Duguet et~al.(2002{\natexlab{a}})]{duguet02a}
Duguet, T., Bonche, P., Heenen, P.-H., and Meyer, J., \emph{Phys. Rev.},
  \textbf{C65}, 014310 (2002{\natexlab{a}}).

\bibitem[Duguet et~al.(2002{\natexlab{b}})]{duguet02b}
Duguet, T., Bonche, P., Heenen, P.-H., and Meyer, J., \emph{Phys. Rev.},
  \textbf{C65}, 014311 (2002{\natexlab{b}}).

\bibitem[Grasso et~al.(2001)]{grasso01a}
Grasso, M., Sandulescu, N., Giai, N.~V., and Liotta, R.~J., \emph{Phys. Rev.},
  \textbf{C64}, 064321 (2001).

\bibitem[Shen et~al.(2003)]{shen03a}
Shen, C., Lombardo, U., Schuck, P., Zuo, W., and Sandulescu, N., \emph{Phys.
  Rev.}, \textbf{C67}, 061302 (2003).

\end{thebibliography}

%%%%%%%%%%%%%%%%%%%%%%%%%%%%%%%%%%%%%%%%%%%
%% Just a reminder that you may have to run bibtex
%% All of it up to \end{document} can be removed
%% if you don't like the warning.
%%%%%%%%%%%%%%%%%%%%%%%%%%%%%%%%%%%%%%%%%%%
\IfFileExists{\jobname.bbl}{}
 {\typeout{}
  \typeout{******************************************}
  \typeout{** Please run "bibtex \jobname" to optain}
  \typeout{** the bibliography and then re-run LaTeX}
  \typeout{** twice to fix the references!}
  \typeout{******************************************}
  \typeout{}
 }

\end{document}